\documentclass[twocolumn,showpacs,prl]{revtex4}
\usepackage{graphics}
\usepackage{epsf}
\usepackage{latexsym}
\usepackage{amsmath}
\usepackage{amssymb}

\def \k{\vec{k}}
\def \grd{^0}

\begin{document}

\author{Bert Reimann$^1$, Reinhard Richter$^1$, Holger Knieling$^1$, Rene Friedrichs$^2$ and Ingo Rehberg$^1$}

\affiliation{$^1$Physikalisches Institut, Experimentalphysik V,
Universit\"at Bayreuth, D-95440 Bayreuth, Germany,}
\affiliation{$^2$FNW/ITP, Otto-von-Guericke-Universit\"at, D-39016 Magdeburg, Germany}

\date{1st October 2004}

\title{Hexagons become second if symmetry is broken}

\begin{abstract}
Pattern formation on the free surface of a magnetic fluid subjected                                          
to a magnetic field is investigated experimentally. 
By tilting the magnetic field the symmetry can be broken in a controllable manner. 
When increasing the amplitude of the tilted field, the flat surface  gives  way 
to liquid ridges. A further increase results in a hysteretic transition to a pattern of
stretched hexagons. The instabilities are detected by means of a linear array of 
magnetic hall sensors and compared with theoretical predictions.
\end{abstract}


\maketitle

Pattern formation in isotropic systems is more complicated than in anisotropic ones:
One of the hallmarks of isotropic systems is the possibility to bifurcate to hexagons 
from an unstructured ground state, which is due to the existence and 
interaction of 3 degenerate wave numbers \cite{cross1993}.
This situation is {\em structurally unstable}, however: 
The smallest distortion of this symmetry acts as a singular perturbation and will lead
to a qualitatively different instability, namely a primary bifurcation to a stripe like pattern.
A specific example has recently been calculated in detail  \cite{friedrichs2002} for a magnetic 
fluid \cite{rosensweig1985}.
In the ideal isotropic system, hexagons will occur under the influence of  a magnetic field 
which is perfectly normal with respect to the fluid surface \cite{cowley1967}.
The slightest change of the orientation 
of the magnetic field is predicted to change this subcritical transition: 
Ridges appear supercritically via the primary 
bifurcation. Their interaction with waves along the less-favoured direction gives rise to  
"stretched" hexagons via a secondary bifurcation. 

 \begin{figure}[hbt]
 \noindent
 \begin{minipage}{8.6cm}
 \begin{center}
 \epsfxsize=8.6cm
 \epsfbox{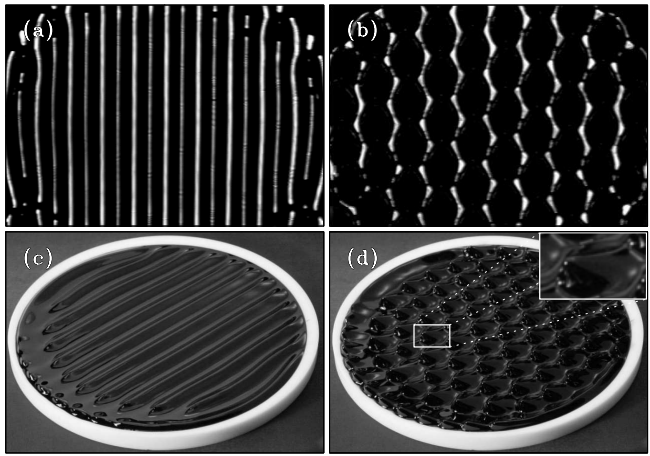}
 \vspace{1mm}
 \caption{Surface patterns of magnetic fluid in a magnetic field tilted by the angle $\varphi = 23^0$ to the
 vertical. Surface reflections of the liquid ridges for the verical induction $\bar{B}=$20 mT (a) and of the tilted crests for
 $\bar{B}=$32 mT (b). The side view of the patterns is presented in (c) and (d), respectively.}
 \label{photos}
 \end{center}
 \end{minipage}
\end{figure}

A first observation of  liquid ridges was reported in Ref. \cite{barkov1977}.
In this paper we present a quantitative characterization of the primary bifurcation to liquid ridges
and a secondary bifurcation to a  pattern of stretched hexagons, as shown in Fig.\ref{photos}, 
via use of a magnetic measurement technique. 
Specifically, we measure the threshold induction $B_{\rm p}$ and $B_{\rm s}$ for the primary 
and secondary instability for various angles of tilt $\varphi$. The measurements 
of $B_{\rm p}( \varphi)$ 
agree with the theoretical prediction if the  nonlinear magnetization curve of the 
magnetic fluid used in the experiment is taken into account. 

\begin{figure}[htb]
\noindent
\begin{minipage}{8.6cm}
\begin{center}
\epsfxsize=7.0cm 
\epsfbox{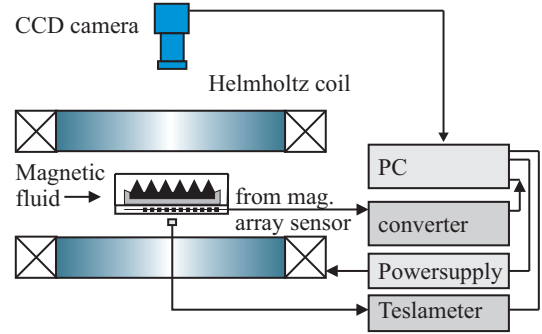}
\vspace{0.1mm}
\caption{Scheme of the experimental setup.}
\label{setup}
\end{center}
\end{minipage}
\end{figure}

Our experimental setup is shown in Fig.\,\ref{setup}. A cylindrical vessel with an edge
machined from Teflon$^{\circledR}$  
with a diameter of 12 cm and a depth of 2 mm is brimful filled with fluid and is situated in the 
center of a pair of Helmholtz coils. For details see Ref.\,\cite{reimann2003}. 
The axis of the coils can be tilted against the vertical by an angle $\varphi=[0^0,90^0]$. 
The experiments are performed with the magnetic fluid 
EMG 909 Lot F061998B (Ferrotec Corp.), with $\mu_r=2.11$. A charge-coupled-device 
(CCD) camera  is recording the patterns from above.  
In order to measure the {\em amplitude} of the steep crests 
a linear array of 32 hall sensors was mounted $1.78\pm 0.1$ mm below the bottom of 
the dish. By this technique the local increase of the magnetic induction below a liquid crest is 
utilized to measure its amplitude. The sensors communicate via 32 amplifiers and a bus with 
the PC. Details of this method are presented elsewere \cite{reimann2003b}. For calibration
purpose  a commercial Hall probe (Group3-LPT-231) in combination with the digital
teslameter (DTM 141) was used. 

\begin{figure}[htb]
  \noindent
  \begin{minipage}{8.6cm}
 \begin{center}
  \epsfxsize=8.6cm

  \epsfbox{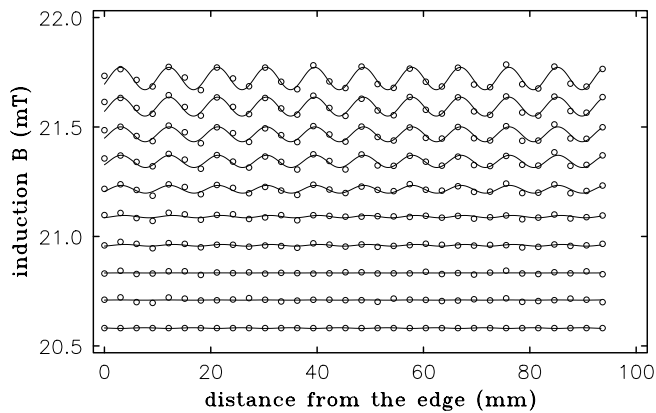}
  \vspace{1mm}
  \caption{Profiles of  $B(x)$, for different  values  of the applied magnetic field.
  The open circles mark the data, the solid lines the fits by 
  Eq.\,(\ref{ridge.amplitude.model}).}
  \label{roll.Profile}
  \end{center}
  \end{minipage}
\end{figure}

The magnetic field is tilted towards the x-axis.
Increasing the magnetic induction, we observe a transition from the flat layer of magnetic fluid to 
the pattern of liquid ridges, displayed in 
Fig.\ref{photos}a,c. The wave vector of the pattern is oriented along the y-axis and 
thus perpendicular to the horizontal field 
component. The vertical component of the local 
magnetic induction was measured by 
means of the sensor array oriented parallel to the wave vector. 
In order to reduce the spatial inhomogeneities of the magnetization caused by the finite 
container size, the spatial variation measured at a subcritical induction  of 20.5 mT is
substracted. The ensuing local induction $B(x)$ is presented in  Fig.\,\ref{roll.Profile} for 
different values of the applied magnetic field, measured at the tilt angle $\varphi= 32^0$. 
The open circles mark the data, the solid line gives the least square fit to
\begin{equation}
B(x)=A\cdot\cos(kx-\psi)+\bar{B}.
\label{ridge.amplitude.model}
\end{equation}
Here $A$ denotes the modulation amplitude, $k$ the absolute value of the wave vector, 
$\psi$ the phase,  and $\bar{B}$ the mean value of the induction.

\begin{figure}[htb]
 \noindent
 \begin{minipage}{8.6cm}
 \begin{center}
 \epsfxsize=8.6cm
 \epsfbox{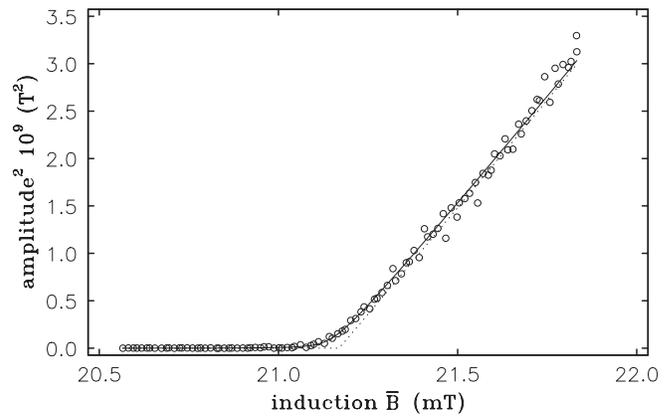}
 \vspace{1mm}
 \caption{Square of the modulation amplitude of the liquid ridges versus the mean magnetic induction $\bar{B}$ of the array 
 detector. For details see text.}
 \label{roll.Amplitude}
 \end{center}
 \end{minipage}
\end{figure}

The square of the modulation amplitude $A$ is plotted 
in Fig.\,\ref{roll.Amplitude} versus the control parameter $\bar{B}$. The 
monotonous increase after a threshold $B_{\rm p}$ is characteristic for a 
supercritical bifurcation. It can be described by the solution of
the stationary amplitude equation  \cite{cross1993}
\begin{equation}
0=\epsilon_{\rm p} A-g A^{3}+b.
\label{ridge.amplitude.stationaer}
\end{equation}
In accordance with the symmetry of the problem
\mbox{$\epsilon_{\rm p}= (\bar{B}^2-B_{\rm p}^2)/B_{\rm p}^2$} was selected to be the 
dimensionless bifurcation parameter. $g$ is the cubic coefficient a scaling-, and $b$ a imperfection parameter. The solid  line in Fig.\ref{roll.Amplitude} gives 
the fit of the experimental data by the solution of  
Eq.(\ref{ridge.amplitude.stationaer}). We obtain 
$B_{\rm p}=21.17 \rm mT$, $g=21.16\rm mT^{-2}$, and 
the slight imperfection $b=4.3 \cdot 10^{-5}\rm mT$.
The dotted line displays the solution without imperfection ($b=0 $mT).

Increasing  the controll parameter further initiates a secondary instability to the stretched 
hexagonal pattern  as displayed in Fig.\ref{photos}b,d. 
The blow--up in (d) indicates that the crests riding on top of 
the ridges are asymmetric with respect to the wave vector of the ridges.
Thus the pattern (of stretched hexagons) lacks any non--trivial rotational symmetry. 

\begin{figure}[htb]
 \noindent
 \begin{minipage}{8.6cm}
 \begin{center}
 \epsfxsize=7.6cm  
 \epsfbox{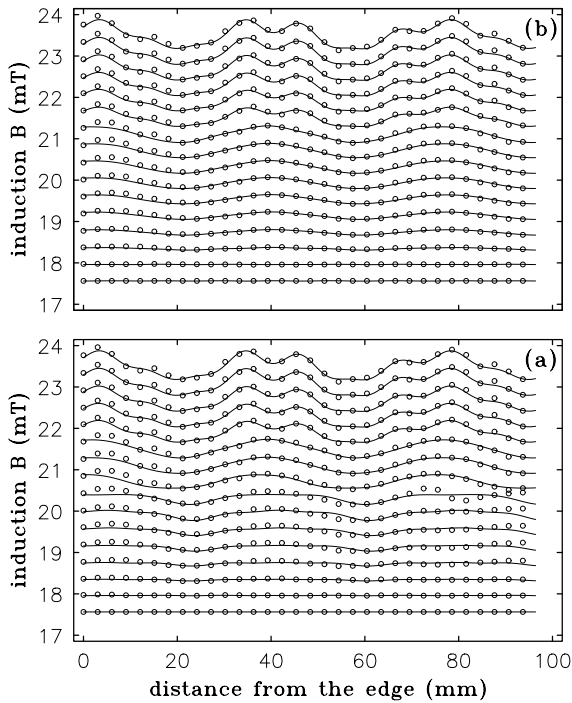}
 \vspace{1mm}
 \caption{ Local magnetic induction for increasing (a) and decreasing (b) control parameter 
 $\bar{B}$ and $\varphi=23^0$. The solid lines give the fit by 
 Eq.\,(\protect\ref{hexarolls.fit.high.harmonisch}).}
  \label{hexaroll.Profile}
 \end{center}
 \end{minipage}
\end{figure}
For a quantitative analysis of the secondary instability a series of 400
measurements of the local induction $B(x)$ has been performed for 
$\varphi=23^0$. For clarity Fig. \ref{hexaroll.Profile}a (b) is presenting only every 20th
line for a quasistatic increase (decrease) of the control parameter $\bar{B}$, 
respectively. In order to detect both, the ridges and the crests, the sensor 
line is now oriented with an angle $\omega=(75.8\pm0.05)\grd$  to the y-axis.
In this way it is covering $2\,\frac{1}{2}$ ridges, which can be 
recognized in the lower part of the plots. For $B\approx22\rm\,mT$ the 
transition to the stretched hexagons occurs. 

A mathematical characterization of the stretched hexagons can be obtained as follows.
In a stretched-hexagonal pattern the wave vectors 
fulfill the side condition $-\vec{k_1}=\vec{k_2}+\vec{k_3}$ and 
$\bar{k}=|\vec{k_2}|=|\vec{k_3}|$. With the abbreviations $k=|\vec{k_1}|$,  
$n=\bar k/k$ and $\tilde b=\sqrt{4n^2-1}$ the wave vectors read
$\vec{k_1}=k(0,1,0)$, $\vec{k_2}=-\frac{k}{2}(-\tilde b, 1, 0)$,
and $\vec{k_3}=-\frac{k}{2}(\tilde b,1,0)$ which coincide for $n=1$ 
with the vectors for a regular hexagonal pattern. The amplitude of 
the ridges $A^R_0$ and of the  stretched-hexagonal pattern  
$A^{H}_0$ can be combined to the amplitude of the 
over-all pattern 
\begin{equation}
A(\vec{x})=A^R_0\cos{\k_1\vec{x}}+\frac{A^{H}_0}{3}\sum_{i=1}^{i=3}{\cos{\k_i\vec{x}}}\;.
\label{hexarolls}
\end{equation}
A cut through this pattern is given by 
\begin{equation}
\vec{x}_M=\vec{x}_0 + t \cdot \vec{e}
\label{line}
\end{equation}
where $t$ denotes the distance from the starting point 
$\vec{x}_0=(x_0,y_0,0)$ of the sensor line and \mbox{
$\vec{e}=(\sin{\omega},\cos{\omega},0)$} its unity vector 
of orientation.
Plugging Eq.(\ref{line}) in Eq.(\ref{hexarolls}) yields 
\begin{equation}
A(\vec{x}_M)=A(t)=A_0^RR(t)+A_0^{H}H(t).
\label{hexarolls.fit.harmonisch}
\end{equation}
for the amplitude along the sensor line. Here 
\begin{equation}
R(t)=\cos{\left(k(y_0+t\cos\omega)\right)}
\end{equation}
gives the contribution of the ridges, and 
\begin{eqnarray}
H(t) & = &\frac{1}{3} \left[ \cos(k(y_0+t\cos\omega)) +\cos(k(\Phi^\star+t\Psi^\star)) \right. \nonumber\\
      & &\left.+\cos(k(\Phi+t\Psi))\right] 
\end{eqnarray}
the contribution of the hexagons. Here
$\Phi=\frac{1}{2}\left(bx_0+y_0\right)\quad,$
$\Psi=\frac{1}{2}\left(b\sin\omega+\cos\omega\right),$
$\Phi^\star=\frac{1}{2}\left(bx_0-y_0\right)\quad$ and
$\Psi^\star=\frac{1}{2}\left(b\sin\omega-\cos\omega\right)$
are abbreviations.
For small amplitudes Eq.(\ref{hexarolls.fit.harmonisch}) is sufficient, 
but for higher amplitudes it is important to take into account the higher harmonics $k_m=m \cdot k$
with $m=1,2,...$ The surface is than given by 
\begin{equation}
A(t)=\sum_{m=1}^{M_R}{A_0^{R_m}R_m(t)}+
     \sum_{m=1}^{M_H}{A_0^{H_{m}}H_m(t)}.
\label{hexarolls.fit.high.harmonisch}
\end{equation}
This model is fitted to the data by four nonlinear parameters, which are
the wave number $k$ of the ridges, the starting point $\vec{x_0}=(x_0,y_0)$
of the sensor line, and the stretching factor $n$ of the hexagonal pattern. 
The amplitudes $A_0^R$ and $A_0^{H}$ are linear parameter of the basic functions 
$R(t)$ and $H(t)$. The solid lines in Fig.\ref{hexaroll.Profile} give the best fit by 
Eq.(\ref{hexarolls.fit.high.harmonisch}) taking into account the basic mode 
of the ridges ($M_R=1$) and the first two   of the hexagons ($M_H=2$). 

\begin{figure}[htb]
 \noindent
 \begin{minipage}{8.6cm}
 \begin{center}
 \epsfxsize=8.6cm
 \epsfbox{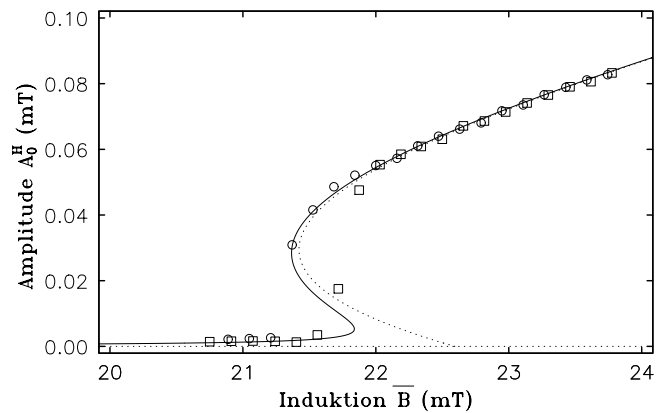}
 \vspace{1mm}
 \caption{Amplitude  $A_0^{\rm H}$ of the stretched hexagonal pattern versus the 
 applied magnetic induction. The open squares 
 (circles) denote the measured amplitude under increase (decrease)
 of the induction, respectively. For clarity only every 10th point is shown. 
 The solid (dotted) line give the fit by Eq.(\protect\ref{hexagon.amplitude.imperfect.eq}) with 
$b= 1.6\cdot10^{-4}\rm mT$ ($b=0 \rm mT$), respectively. 
For the other parameters we obtained $B_{\rm s}= 22.594\rm mT$, $\gamma_1=7.6 \rm mT^{-1}$, $\gamma_2=0.9 \rm mT^{-1}$,
and $g=116.31 \rm mT^{-2}$. 
}
 \label{hexaroll.Amplitude}
 \end{center}
 \end{minipage}
\end{figure}

From this fit the amplitude $A_0^{H}$ of the hexagons can be extracted. It is plotted in 
Fig.\ref{hexaroll.Amplitude} vs. the control parameter $\bar{B}$. The open squares (circles) 
mark the data for an increase (decrease) of $\bar{B}$, respectively. 
The hysteresis is characteristic for a subcritical bifurcation,
which has been predicted for the transition from ridges to stretched hexagons \cite{friedrichs2002}.

Next we describe the amplitude \mbox{$A_H \equiv A^H_0$} of the hexagons after the secondary 
bifurcation at 
\mbox{$\epsilon_{\rm S}= (\bar{B}^2-B_{\rm S}^2)/B_{\rm P}^2$}.
In the spirit of a weakly nonlinear analysis slightly above
$\epsilon_{\rm S}$ we use the amplitude equation
\begin{equation}
0=\epsilon A_H+\gamma_1(1+\gamma_2 \epsilon_S)A_H^2-g A_H^3+b_S.
\label{hexagon.amplitude.imperfect.eq}
\end{equation}
In this experimental paper the 
coefficients in Eq.(\ref{hexagon.amplitude.imperfect.eq}) have been obtained 
by a fit to the measurements in order to circumvent their tedious calculation from the 
basic equations. To avoid the ambiguity of $A_H(\bar{B})$ in the 
hysteretic regime, $\bar{B}(A_H)$ was fitted to the data according to Ref. \cite{aitta1985}.
The result of the fit is presented in Fig.\ref{hexaroll.Amplitude} by a solid line,
while the dashed line gives the solution for the same parameters, however with  $b_S=0$. 

For decreasing $\bar B$ the system follows the solid line very well down to 
the saddle-node. For increasing $\bar B$ the agreement is less convincing in the bistable regime.
Here the impact of the edges \cite{pfister1981} seems to penetrate the interior of the dish much 
stronger. As a consequence the analysis by Eq.(\ref{hexagon.amplitude.imperfect.eq}) is 
not sufficient in this regime - see also Fig.\ref{hexaroll.Profile}a. 

\begin{figure}[htb]
 \noindent
 \begin{minipage}{8.6cm}
 \begin{center}
 \epsfxsize=8.6cm
 \epsfbox{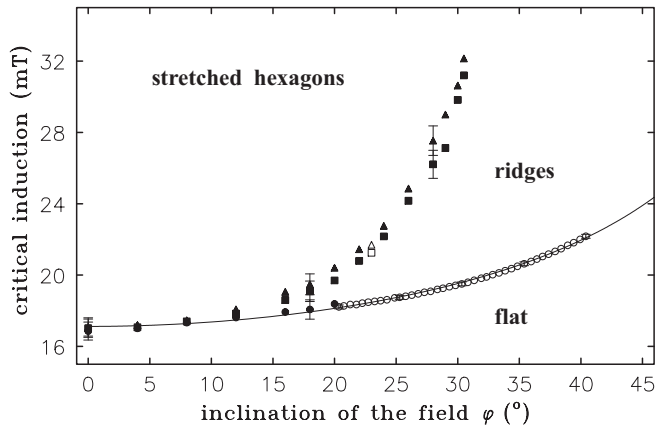}
 \vspace{1mm}
 \caption{Critical inductions versus the inclination of the magnetic field.
  The data marked by open circles have been estimated by 
 fitting the evolution of the ridge amplitude (e.g. Fig.\,\protect\ref{roll.Amplitude})  by 
 Eq. (\protect\ref{ridge.amplitude.stationaer}). The other values, marked by full symbols, have been determined 
by visual inspection of the liquid layer.  The solid line gives the fit obtained via
 Eq.\,(\protect\ref{dispersion.relation}).}
 \label{crit.induction}
 \end{center}
 \end{minipage}
\end{figure}
Next we investigate the angular dependence of the critical induction for
the first and secondary bifurcation. In Fig.\,\ref{crit.induction} the measured data for
the transition to ridges are marked by circles, whereas the transition to 
stretched hexagons is denoted by triangles and the reverse transition to ridges by squares. 

The solid line gives $B_{\rm p}$ calculated for the instability of the flat
surface. It follows from the dispersion relation of the surface waves in $y$-direction
with $\omega^2=0$ and $k_y=k_c=\sqrt{\rho g/\sigma}$,
using the relation between magnetic induction and magnetization,
\begin{eqnarray}
\label{dispersion.relation}
\omega^2(k_x=0,k_y)&=&g|k_y|+\frac{\sigma}{\rho}|k_y|^3-\frac{\mu_0}{\rho}k_y^2M_Z^2\\
&\times &
\frac{M^2-M_Z^2\frac{\bar{\chi}-\chiï}{\bar{\chi}+1}}{M^2-M_Z^2\frac{\bar{\chi}-\chiï}{\bar{\chi}+1}
+\frac{M}{\bar{\chi}+1}\sqrt{M^2-M_Z^2\frac{\bar{\chi}-\chiï}{\bar{\chi}+1}}}\nonumber.
\end{eqnarray}
The dispersion relation Eq.(\ref{dispersion.relation}) takes into account
the nonlinearity of the magnetization
curve $M(H)$ of the fluid and can be deduced from Eq.(36) in
\cite{zelazo1969} whereby ${\bf M}=M_Z{\bf e}_{z} + M_X{\bf e}_{x}$ denotes
the magnetization of the fluid for the undisturbed surface.
The susceptibilities ${\bar \chi}(H) = \frac{M(H)}{H}$ and
${\chi'}(H) = \frac{\partial M(H)}{\partial H}$ were determined from
the experimental magnetization curve assuming a logarithmic normal
distribution for the size of the magnetic particles in the fluid
\cite{embs2001,friedrichs2003}. In contrast, in Ref.\,\cite{friedrichs2002} a constant $\chi$ has 
been used, which results in a $B_{\rm p}$ not depending on $\varphi$.

To conclude, for the tilted field instability we have measured the forward bifurcation to liquid 
ridges. The angular dependence observed in the experiment, is quantitatively 
described by taking into account the nonlinear magnetization. 
In addition we measured the backward bifurcation 
to hexagons, which has been predicted by an  energy variational method\,\cite{friedrichs2002}.
A full quantitative agreement with these predictions can not be expected,
because the theory is restricted to permeabilities $\mu_r<1.4$, while we had to use 
$\mu_r=2.11$ to avoid huge fields. 
The essence of the experimental observation, namely a structural change of the primary 
instability, seems to be well described by this theoretical ansatz:
For broken symmetry ridges always precede hexagons. They are  increasingly difficult 
to resolve, however, if the angle of tilt diminishes.
A similar scenario can be expected e.g., for non-boussinesq inclined layer convection 
\cite{bodenschatz2000}, for  magnetohydrodynamic  as well as electro-convection in tilted 
magnetic fields \cite{busse1990},
for lucent hexagons under influence of an asymmetric Fourier filter \cite{ackemann1999},
and for Turing patterns \cite{ouyang1991} in stressed gel.
 
\begin{acknowledgments}
We thank W.Pesch for fruitful discussions and {\sl Deutsche Forschungsgemeinschaft} for financial support under grant Ri 1054/1-3.
\end{acknowledgments}


\begin{thebibliography}{10}

\bibitem{cross1993}
M.~C. Cross and P.~C. Hohenberg, Rev. Mod. Phys. {\bf 65},  870  (1993).

\bibitem{friedrichs2002}
R. Friedrichs, Phys. Rev. E {\bf 66},  066215  (2002).

\bibitem{rosensweig1985}
R.~E. Rosensweig, {\em Ferrohydrodynamics} (Cambridge University Press,
  Cambridge, 1985).

\bibitem{cowley1967}
M.~D. Cowley and R.~E. Rosensweig, J. Fluid Mech. {\bf 30},  671  (1967);
A. Gailitis, J. Fluid Mech. {\bf 82},  401  (1977).

\bibitem{barkov1977}
Y.~D. Barkov and V.~G. Bashtovoi, Magnetohydrodynamics (N.Y.) {\bf 13},  497
  (1977).

\bibitem{reimann2003}
B. Reimann, R. Richter, I. Rehberg, and A. Lange, Phys. Rev. E {\bf 68},
  036220  (2003).

\bibitem{reimann2003b}
B. Reimann, {\em Experimente zur Normal- und Schr\"ag\-feld\-instabilit\"at
  magnetischer Fl\"ussig\-keiten}, (Shaker Verlag, Aachen, 2003).

\bibitem{friedrichs2001}
R. Friedrichs and A. Engel, Phys. Rev. E {\bf 64},  021406  (2001).

\bibitem{aitta1985}
A. Aitta, G. Ahlers, and D.~S. Cannel, Phys. Rev. Lett. {\bf 54},  673  (1985).

\bibitem{pfister1981}
G. Pfister and I. Rehberg, Physics Letters {\bf 83A},  19  (1981).

\bibitem{zelazo1969}
R.~E. Zelazo and J.~R. Melcher, Fluid Mech {\bf 39},  1  (1969).

\bibitem{friedrichs2003}
R. Friedrichs, Ph.D. thesis, Universit\"at Magdeburg, 2003.

\bibitem{embs2001}
J. Embs et al.,
Magnetohydrodynamics {\bf 37},  222  (2001).

\bibitem{bodenschatz2000}
K.~E. Daniels, B.~B. Plapp, and E. Bodenschatz, Phys. Rev. Lett. {\bf 84},
  5320  (2000).

\bibitem{busse1990}
F.~H. Busse and R.~M. Clever, Eur. J. Mechanics B {\bf 9},  225  (1990).
A. Buka, B. Dressel, L. Kramer, and W. Pesch, Phys. Rev. Lett. {\bf 93},
044502  (2004).

\bibitem{ouyang1991}
Q. Ouyang and H. Swinney, Nature {\bf 352},  610  (1991).

\bibitem{ackemann1999}
T. Ackemann, B. Giese, B. Sch\"apers, and W. Lange, J. Opt. B: Quantum
  Semiclassical Opt. {\bf 1},  70  (1999).

\end{thebibliography}
\end{document}